\documentclass[twocolumn,pra,aps,floatfix,longbibliography]{revtex4-1}
\usepackage{amsmath}
\usepackage{amssymb}
\usepackage{mathrsfs}
\usepackage{graphicx}
\usepackage{lipsum}
\usepackage{dsfont}
\usepackage{comment}
\usepackage{appendix}
\usepackage{xfrac}
\begin{document}
\title{Unidirectional Maxwellian Spin Waves}
\author{Todd Van Mechelen}
\author{Zubin Jacob}
\email{zjacob@purdue.edu} 
\affiliation{Purdue University, School of Electrical and Computer Engineering, Brick Nanotechnology Center, 47907, West Lafayette, Indiana, U.S.A.}

\begin{abstract}
\noindent
We develop a unified perspective of unidirectional topological edge waves in non-reciprocal media. We focus on the inherent role of photonic spin in non-reciprocal gyroelectric media, ie. magnetized metals or magnetized insulators. Due to the large body of contradicting literature, we point out at the outset that these Maxwellian spin waves are fundamentally different from well-known topologically trivial surface plasmon polaritons (SPPs). We first review the concept of a Maxwell Hamiltonian in non-reciprocal media, which immediately reveals that the gyrotropic coefficient behaves as a photon mass in two dimensions. Similar to the Dirac mass, this photonic mass opens bandgaps in the energy dispersion of bulk propagating waves. Within these bulk photonic bandgaps, three distinct classes of Maxwellian edge waves exist - each arising from subtle differences in boundary conditions. On one hand, the edge wave solutions are rigorous photonic analogs of Jackiw-Rebbi electronic edge states. On the other hand, for the exact same system, they can be high frequency photonic counterparts of the integer quantum Hall effect, familiar at zero frequency. Our Hamiltonian approach also predicts the existence of a third distinct class of Maxwellian edge wave exhibiting topological protection. This occurs in an intriguing topological bosonic phase of matter, fundamentally different from any known electronic or photonic medium. The Maxwellian edge state in this unique \textit{quantum gyroelectric phase of matter} necessarily requires a sign change in gyrotropy arising from non-locality (spatial dispersion). In a Drude system, this behavior emerges from a spatially dispersive cyclotron frequency that switches sign with momentum. A signature property of these topological electromagnetic edge states is that they are oblivious to the contacting medium, ie. they occur at the interface of the quantum gyroelectric phase and any medium (even vacuum). This is because the edge state satisfies open boundary conditions - all components of the electromagnetic field vanish at the interface. Furthermore, the Maxwellian spin waves exhibit photonic spin-1 quantization in exact analogy with their supersymmetric spin-\sfrac{1}{2} counterparts. The goal of this paper is to discuss these three foundational classes of edge waves in a unified perspective while providing in-depth derivations, taking into account non-locality and various boundary conditions. Our work sheds light on the important role of photonic spin in condensed matter systems, where this definition of spin is also translatable to topological photonic crystals and metamaterials.
\end{abstract}
\maketitle
\noindent
\noindent
\vspace{-30 pt}
\section{Introduction}

Gyroelectric media, or magnetized plasmas, form the canonical system to study non-reciprocity \cite{Mackay:16,Caloz2018,Mirmoosa2014,Valente2015,Floess_2018,Zyuzin2017}. There has been recent interest in such media for their potential to break the time-bandwidth limit inside cavities  \cite{Mann:19,Tsakmakidis1260}, sub-diffraction imaging \cite{Zhang2011}, unique absorption \cite{Green2012} and thermal properties \cite{Zhu2016}, and for one-way topological transitions \cite{Leviyev2017}. It should be emphasized that the gyroelectric coefficient ($g$), which embodies antisymmetric components of the permittivity tensor ($\varepsilon_{ij}$), is intimately related to its low frequency counterpart in condensed matter physics - the transverse Hall conductivity ($\sigma_{H}=\sigma_{xy}=-i\omega g$) \cite{STERN2008204,Stefaan2012}. The goal of this paper is to bridge the gap between modern concepts in nanophotonics, magnetized plasma physics, and condensed matter physics.

Historically, gyroelectric media was popularized in plasma physics \cite{landau1975classical,landau2013electrodynamics} where the ``gyration vector" or ``rotation axis" sets a preferred handedness to the medium. This causes non-reciprocal (direction dependent) wave propagation along the axis of the medium. The non-reciprocal properties are now well understood but only recently has the connection with the Dirac equation been revealed \cite{van2018dirac,VanMechelen2018,VanMechelen:19,Horsley2018,Bialynicki_Birula_2013,Barnett_2014,Horsley2018}. This immediately leads to multiple new insights related to energy density, photon spin and photon mass for wave propagation within two-dimensional gyrotropic media \cite{VanMechelen2018,VanMechelen:19,Horsley2018,Dunne1999}. In particular, a unique phenomenon related to gyrotropic media is the presence of unidirectional edge waves, fundamentally different from surface plasmon polaritons (SPPs) or Dyakonov waves \cite{Filipa2018,Davoyan}. We note that photonic crystals \cite{Lu2016,Shalaev_2018,Noh2018} or metamaterials \cite{Khanikaev2017,Chang2017,Lineaat2774} are not necessary for this phenomenon and even a continuous medium (eg: magnetized plasma or doped semiconductor) can host unidirectional edge waves.

The role of spin has not been revealed till date but chiral (unidirectional) photonic waves in gyrotropic media have a rich history. Early work introduced the concept of optical isomers \cite{Zhukov2000} which is the interface of two gyrotropic media with opposite signs of non-reciprocal coefficients (half-space of $g>0$ interfaced with another half-space of $g<0$). It was shown that unique chiral edge states emerge, addressed as the ``quantum Cotton-Mouton effect", which are similar in nature to the electronic quantum Hall effect. These chiral edge states were also predicted on the interface of Weyl semimetals \cite{Zyuzin2015}. Raghu and Haldane's original model to realize a one-way waveguide dealt with the gyroelectric photonic crystals \cite{Haldane2008,Raghu2008}. More recently, gyroelectric magneto-plasmons have been demonstrated in quantum well structures under biasing magnetic fields \cite{Jin2016,Manfra2017}. Another important example of unidirectional edge waves occurs when a gyrotropic medium is terminated with a perfect electric conductor (PEC), as shown by Silveirinha \cite{Lannebere2018}. Horsley \cite{Horsley2018} recently proved that this PEC boundary is equivalent to antisymmetric solutions of optical isomers (two gyrotropic media with opposite signs $\pm g$) and leads to unidirectional Jackiw-Rebbi type photonic waves.

However, in all the above examples, the electromagnetic boundary conditions are drastically different from the \textit{open} boundary conditions utilized for topologically-protected solutions of the Dirac equation \cite{Delplace2011,Mong2011,Shen2011,shen2017topological,Medhi_2012,bernevig2013topological}. This challenge was recently overcome when a Dirac-Maxwell correspondence was applied to gyrotropic media \cite{VanMechelen2018,VanMechelen:19}, which derived the supersymmetric (spin-1) partner of the topological Dirac equation. This framework gave rise to a new unidirectional edge wave with open boundary conditions, such that the electromagnetic field completely vanishes at the material interface \cite{VanMechelen2018,VanMechelen:19}. The necessary conditions for the existence of such a wave is non-reciprocity $g$, temporal dispersion $g(\omega)$, and spatial dispersion $g(\omega,k)$. A momentum dependent sign change in the gyrotropic coefficient $g(\omega,k_\mathrm{crit})=0$ leads to a topologically nontrivial electromagnetic field - a quantum gyroelectric phase of matter. In Drude systems, this corresponds to a momentum dependent sign change of the cyclotron frequency. It should be emphasized that this topological phase of matter is Maxwellian (spin-1 bosonic) and is unlike any known spin-\sfrac{1}{2} fermionic phases of matter (eg: graphene, Chern insulator, etc.). The unidirectional photonic edge wave is a fundamental mode of this nonlocal, non-reciprocal medium and cannot be separated from the bulk. The contacting medium has no influence on the edge wave, unlike the previously mentioned examples which are sensitive to boundary conditions. We address this phenomenon as the quantum gyroelectric effect (QGEE) and it remains an open question whether such a Maxwellian phase of matter can be found in nature.

\begin{figure*}
\includegraphics[width=\linewidth]{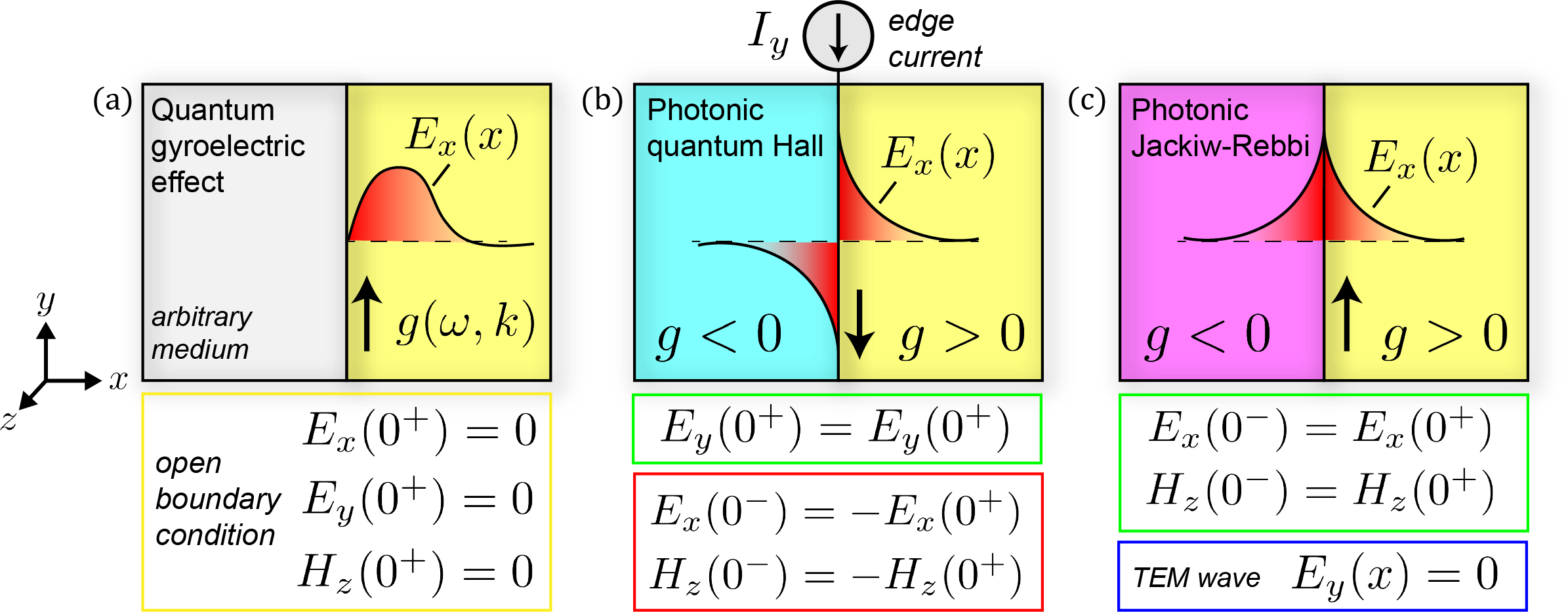}
\caption{(a), (b) and (c) are schematics of the quantum gyroelectric effect (QGEE), photonic quantum Hall (PQH) and photonic Jackiw-Rebbi (PJR) edge states respectively. The characteristic spatial profile of $E_x(x)$ is displayed for each edge state along with the corresponding boundary conditions. (a) The QGEE is a topologically-protected unidrectional (chiral) edge state and exists at the boundary of any medium - even vacuum. The QGEE is fundamentally tied to nonlocal (spatially dispersive) gyrotropy $g(\omega,k)$ and can never be realized in a purely local model. (b) The PQH edge state is the photonic analogue of the quantum Hall effect and hosts a high-frequency edge current $I_y$. The presence of the edge current $I_y\neq 0$ creates a discontinuity in the fields across the boundary, $E_x(0^-)\neq E_x(0^+)$ and $H_z(0^-)\neq H_z(0^+)$. (c) The PJR edge state is the photonic equivalent of the inverted mass problem arising in the Dirac equation. This state possesses no edge current $I_y= 0$ and is completely transverse electro-magnetic (TEM) as the longitudinal field vanishes entirely $E_y(x)=0$.}
\label{fig:EdgeStates}
\end{figure*}

The purpose of this paper is to present the first unified view of all the aforementioned unidirectional edge waves in non-reciprocal media. The essence of our results is captured in Fig.~\ref{fig:EdgeStates} and Tab.~\ref{tab:SummEdgeStates} which contrasts unidirectional edge waves of the quantum gyroelectric effect (QGEE), photonic quantum Hall (PQH) states and photonic Jackiw-Rebbi (PJR) states. All such waves appear in gyroelectric media but boast surprisingly different behavior. The QGEE displays bulk-boundary correspondence \cite{Mong2011} since it is defined independent of the contacting medium [Sec.~\ref{sec:QGEE}]. The PQH states host a high frequency quantum Hall edge current which arises from a discontinuity in the electromagnetic field [Sec.~\ref{sec:QuantumHall}]. Lastly, the PJR edge waves are domain wall states [Sec.~\ref{sec:JackiwRebbi}]. Another important result of our paper is illustrated in Fig.~\ref{fig:MirrorBC} which shows that the two classes of unidirectional waves, PQH and PJR, can be realized at perfect magnetic conductor (PMC) and perfect electric conductor (PEC) boundary conditions respectively.

This article is organized as follows. Sec.~\ref{sec:Overview} presents an overview of spin waves. In Sec.~\ref{sec:MaxwellHamiltonain} and \ref{sec:QGEE} we show that a nonlocal, non-reciprocal medium is foundational to the concept of 2+1D topological phases of matter. We review the concept of Dirac-Maxwell correspondence which can be exploited to introduce a Hamiltonian for light within complex photonic media. This framework allows us to rigorously define helicity and spin while also identifying a photonic mass, which is directly proportional to the gyrotropic coefficient. We then discuss the necessity of temporally and spatially dispersive optical response parameters to define electromagnetic topological invariants for bulk continuous media. Although commonly ignored, nonlocality is absolutely essential for the electromagnetic theory to be consistent with the tenfold way \cite{Ryu_2010}, which describes all possible continuum topological phases, in every dimension. In the topologically nontrivial regime $C\neq 0$, the unidirectional Maxwellian spin wave is derived and satisfies open boundary conditions - this is the QGEE. Following these results, we analyze the interface of optical isomers [Sec.~\ref{sec:OpticalIsomers}] and derive the photonic quantum Hall [Sec.~\ref{sec:QuantumHall}] and photonic Jackiw-Rebbi edge states [Sec.~\ref{sec:JackiwRebbi}]. The final Sec.~\ref{sec:Conclusions} presents our conclusions.

\section{Overview of spin waves}\label{sec:Overview}

We outline the key properties of chiral Maxwellian spin waves which, surprisingly, emerge in two distinct physical systems. First, it is identified in the low momentum dispersion of the QGEE. Second, it also represents the photonic counterpart of the Jackiw-Rebbi domain wall state known in the continuum Dirac equation \cite{Shen2011,shen2017topological,Jackiw1976,Schuster2016}. The Dirac Jackiw-Rebbi wave exists at the interface of inverted masses, $\Lambda>0$ and $\Lambda<0$, and is an \textit{eigenstate} of the spin-\sfrac{1}{2} helicity (Pauli) operator. The exact parallel in photonics can now be established as it has been proven that gyrotropy plays the role of photonic mass. Thus, a unique Maxwellian spin wave exists at the interface of optical isomers, $g>0$ and $g<0$. Furthermore, this electromagnetic wave is an eigenstate of the SO(3) operator (spin-1 helicity operator) and exhibits \textit{helical quantization}. This is intuitively clear since the edge wave is purely transverse electro-magnetic (TEM); the polarization is orthogonal to the momentum $\Hat{k}\cdot\vec{E}=0$.

To avoid confusion, we contrast between conventional surface plasmon polaritons (SPPs) and Maxwellian spin waves which both display spin-momentum locking phenomena but in fundamentally different forms. Even SPPs on magnetized plasmas do not show the same characteristics as chiral Maxwellian spin waves as they are not eigenstates of the SO(3) vector operators. We strongly emphasize that SPPs on conventional (electric) metals, magnetic metals, as well as negative index media \cite{Smith788} do not possess any topological characteristics. There exists no bulk-boundary correspondence as the bulk media are trivial. Spin-momentum locking in these {\color{red}surface} waves is transverse and \textit{not} quantized \cite{VanMechelen:16,Farid2016,Bliokh1448,Mitsch2014,Young2015,Bliokh2015,Lodahl2017,Picardi2018}. This means the spin is perpendicular to the momentum and is a continuous (classical) number. On the contrary, spin-momentum locking arising in Maxwellian spin waves is \textit{longitudinal} and \textit{quantized}. This means the spin is parallel to the momentum and is a discrete (quantum) number, assuming values of $\pm 1$ only. Despite recent observations of spin-momentum locking phenomena in waveguides \cite{Kapitanova2014,Luo2017}, resonators \cite{Wang2019} and surface plasmon polaritons \cite{Lin331}, no wave has been discovered to be a pure spin state with quantized eigenvalues of the helicity operator. Our work is an answer to this endeavor. 

As an aside, we must also point out that orbital angular momentum (OAM) quantization for photons is unrelated to topological quantization \cite{carroll2004spacetime}, such as Chern number quantization. OAM quantization is routinely encountered for classical optical waves in free-space beams \cite{Gawhary2018}, microdisk resonators, optical fibers, whispering gallery mode resonators \cite{Yao:11}, etc. The origin of \textit{topological} quantization is always a singularity/discontinuity in the underlying gauge potential \cite{WU1976365,Arttu2012,Fang92}. This phenomenon of gauge singularity/discontinuity has been proven to occur in the Berry connection of the quantum gyroelectric phase \cite{VanMechelen2018,VanMechelen:19}. Nevertheless, it remains an open question whether such topological quantization is connected to physical observables (response/correlation functions) of the photon, like they are for the electron. For example, quantization of the Hall conductivity $\sigma_H$ was the first striking experimental observable connected to topology \cite{Hatsugai1993,Hatsugai_1997}. No photonic equivalent is known to date.



 



\section{Maxwell Hamiltonian}\label{sec:MaxwellHamiltonain}

\subsection{Vacuum}

Before defining Maxwellian spin waves [Fig.~\ref{fig:EdgeStates}], that emerge at the boundaries of matter, we illustrate the direct correspondence of spin operators arising in Maxwell's equations and the massless Dirac equation in 2+1D. We will then show that this correspondence extends to \textit{massive} particles in Sec.~\ref{subsec:PhotonMass}. In two spatial dimensions we can focus strictly on transverse-magnetic (TM) waves, where the magnetic field $H_z$ is perpendicular to the plane of propagation $\mathbf{k}=k_x\hat{x}+k_y\hat{y}$. Maxwell's equations in the reciprocal momentum space $\mathcal{H}_0(\mathbf{k})$ are expressed compactly as \cite{VanMechelen2018,VanMechelen:19,Horsley2018},
\begin{equation}\label{eq:VacuumMaxwell}
\mathcal{H}_0(\mathbf{k}) f=\omega f, \qquad f=\begin{bmatrix}
E_x\\E_y\\H_z
\end{bmatrix}.
\end{equation}
$f$ is the TM polarization of the electromagnetic field and is operated on by the free-space ``Maxwell Hamiltonian",
\begin{equation}
\mathcal{H}_0(\mathbf{k})=\begin{bmatrix}
0 & 0 &-k_y\\
0 & 0 & k_x\\
-k_y & k_x & 0
\end{bmatrix}=k_x\hat{S}_x+k_y\hat{S}_y.
\end{equation}
Maxwell's equations describe optical helicity, ie. the projection of the momentum $\mathbf{k}$ onto the spin $\vec{S}$. In this case, $\hat{S}_x$ and $\hat{S}_y$ are spin-1 operators that satisfy the angular momentum algebra $[\hat{S}_i,\hat{S}_j]=i\epsilon_{ijk}\hat{S}_k$. These operators are expressed in matrix form as,
\begin{equation}\label{eq:Spin1Operators}
\hat{S}_x=\begin{bmatrix}
0 & 0 & 0\\
0 & 0 & 1\\
0 & 1 & 0
\end{bmatrix},~~~ \hat{S}_y=\begin{bmatrix}
0 & 0 & -1\\
0 & 0 & 0\\
-1 & 0 & 0
\end{bmatrix}, ~~~ \hat{S}_z=\begin{bmatrix}
0 & -i & 0\\
i & 0 & 0\\
0 & 0 & 0
\end{bmatrix}.
\end{equation}
$\hat{S}_z$ is the spin-1 operator along $\hat{z}$ and generates rotations in the $x$-$y$ plane. As we will see, $\hat{S}_z$ is fundamentally tied to photonic mass in two dimensions. To prove this, we will first review the definition of mass for two-dimensional Dirac particles and show there is a one-to-one correspondence with photons.

\subsection{Dirac equation}

For comparison, consider the two-dimensional massless Dirac equation, which often describes the quasiparticle dynamics of graphene \cite{Pal2011,gu2011relativistic,MOHAN2018200}. This is also known as the Weyl equation,
\begin{equation}
H_0(\mathbf{k})\Psi=E\Psi.
\end{equation}
$\Psi$ is a two-component spinor function and is acted on by the massless Dirac Hamiltonian,
\begin{equation}\label{eq:MasslessDirac}
H_0(\mathbf{k})=k_x\sigma_x+k_y\sigma_y.
\end{equation}
Like Maxwell's equations, the Weyl equation represents electronic helicity - the projection of momentum $\mathbf{k}$ onto the spin $\vec{\sigma}$. In this case, $[\sigma_i,\sigma_j]=2i\epsilon_{ijk}\sigma_k$ are the Pauli matrices and describe the dynamics of a spin-\sfrac{1}{2} or pseudospin-\sfrac{1}{2} particle,
\begin{equation}
\sigma_x=\begin{bmatrix}
0 & 1\\
1 &0
\end{bmatrix}, \qquad \sigma_y=\begin{bmatrix}
0 & -i\\
i & 0
\end{bmatrix}, \qquad \sigma_z=\begin{bmatrix}
1 & 0\\
0 & -1
\end{bmatrix}.
\end{equation}
As we can see, the $\sigma_z$ Pauli matrix is clearly missing from the Weyl equation [Eq.~(\ref{eq:MasslessDirac})]. We cannot add a term proportional to $\sigma_z$ due to time-reversal symmetry,
\begin{equation}
\mathcal{T}^{-1}H_0(-\mathbf{k})\mathcal{T}=H_0(\mathbf{k}), \qquad \mathcal{T}=i\sigma_y\mathcal{K}.
\end{equation}
$\mathcal{K}$ represents the complex conjugation operator in this context and $\mathcal{T}^2=-\mathds{1}_2$ is a fermionic operator. 

However, if we \textit{break} time-reversal symmetry $\mathcal{T}^{-1}H(-\mathbf{k})\mathcal{T}\neq H(\mathbf{k})$ then $\sigma_z$ is permitted. This transforms the massless Weyl equation to the \textit{massive} Dirac equation $H_0(\mathbf{k})\to H(\mathbf{k})$,
\begin{equation}\label{eq:DiracMass}
H(\mathbf{k})=v(k_x\sigma_x+k_y\sigma_y)+\Lambda\sigma_z.
\end{equation}
We have also introduced the Fermi velocity $v$ which describes the effective electron speed. Equation~(\ref{eq:DiracMass}) models a multitude of problems in condensed matter physics, such as Dirac particles and the $p+ip$ wave superconductor \cite{Read2000}. The Dirac mass $\Lambda$ has many important properties. It respects rotational symmetry in the $x$-$y$ plane and opens a band gap at $E=0$,
\begin{equation}\label{eq:DispersionDirac}
E^2-\Lambda^2=v^2k^2,
\end{equation}
with $k^2=k_x^2+k_y^2$. It is clear that when $E^2<\Lambda^2$, waves decay exponentially into the medium. The rest energy $E^2=\Lambda^2$ defines the stationary point $k=0$. Furthermore, the Dirac mass also breaks parity (mirror) symmetry in both $x$ and $y$ dimensions. For Dirac particles, the mirror operators are simply,
\begin{equation}
\mathcal{P}_x=\sigma_y, \qquad \mathcal{P}_y=\sigma_x.
\end{equation}
One can easily check that $\mathcal{P}_x^{-1} H(-k_x)\mathcal{P}_x\neq H(k_x)$ and $\mathcal{P}_y^{-1} H(-k_y)\mathcal{P}_y\neq H(k_y)$ do not commute when $\Lambda\neq 0$. A review of Jackiw-Rebbi Dirac states arising at the interface of inverted masses $\pm\Lambda$ is presented in Appendix \ref{app:JRDirac}.

\subsection{Definition of photon mass in gyrotropic media}\label{subsec:PhotonMass}

The question now: what is the equivalent of mass for the photon? In analogy with the Dirac equation, the photon mass must respect rotational symmetry but break parity and time-reversal. The answer is a bit subtle. There are two components of the permittivity tensor $\varepsilon_{ij}$ that are permitted by rotational symmetry in the plane,
\begin{equation}
\varepsilon_{ij}=\varepsilon\delta_{ij}+ig\epsilon_{ij}.
\end{equation}
$\varepsilon$ is the diagonal part (scalar permittivity) and $g$ is the off-diagonal part (gyrotropy). $\epsilon_{ij}=-\epsilon_{ji}$ is the 2D antisymmetric tensor and should not be confused with the permittivity tensor $\varepsilon_{ij}$ itself. To put Maxwell's equations into a more enlightening form, we normalize $f$ by,
\begin{equation}
f\to \mathfrak{F}=\begin{bmatrix}
\sqrt{\varepsilon}E_x\\
\sqrt{\varepsilon}E_y\\
H_z
\end{bmatrix}.
\end{equation}
Inserting the permittivity tensor, the vacuum wave equation [Eq.~(\ref{eq:VacuumMaxwell})] is transformed to $\mathcal{H}_0(\mathbf{k})\to\mathcal{H}(\mathbf{k})$,
\begin{equation}
\mathcal{H}(\mathbf{k})\mathfrak{F}=\omega\mathfrak{F},
\end{equation}
where the effective Maxwell Hamiltonian is expressed as,
\begin{equation}\label{eq:MassiveMaxwell}
\mathcal{H}(\mathbf{k})=v_p(k_x\Hat{S}_x+k_y\Hat{S}_y)+\Lambda_p\Hat{S}_z.
\end{equation}
By direct comparison with the massive Dirac equation [Eq.~(\ref{eq:DiracMass})], we see that $v_p$ is the effective speed of light and $\Lambda_p$ is the effective photon mass,
\begin{equation}
v_p=\frac{1}{\sqrt{\varepsilon}}, \qquad \Lambda_p=\omega\frac{g}{\varepsilon}.
\end{equation}
The one significant difference between the two equations is that $\vec{S}$ are spin-1 operators while $\vec{\sigma}$ are spin-\sfrac{1}{2} operators. This is intuitive because the photon is a bosonic particle. In fact, massive Dirac particles [Eq.~(\ref{eq:DiracMass})] and massive photons [Eq.~(\ref{eq:MassiveMaxwell})] are \textit{supersymmetric partners} in two dimensions \cite{Dunne1999,Stephen2001}. It should be emphasized however, that $\varepsilon$ and $g$ are always dispersive which means the effective speed $v_p=v_p(\omega)$ and effective mass $\Lambda_p=\Lambda_p(\omega)$ depend on the energy $\omega$.


Like the Dirac equation, the photon mass $\Lambda_p\neq 0$ is proportional to the $\Hat{S}_z$ operator and breaks time-reversal symmetry,
\begin{equation}
\mathcal{T}^{-1}\mathcal{H}(-\mathbf{k})\mathcal{T}\neq \mathcal{H}(\mathbf{k}), \qquad \mathcal{T}=\begin{bmatrix}
1 & 0 & 0\\
0 & 1 & 0\\
0 & 0 & -1
\end{bmatrix}\mathcal{K},
\end{equation}
where $\mathcal{T}^2=+\mathds{1}_3$ is a bosonic operator. For photons, the mirror operators in the $x$ and $y$ dimensions are defined as,
\begin{equation}
\mathcal{P}_x=\begin{bmatrix}
-1 & 0 & 0\\
0 & 1 & 0\\
0 & 0 & -1
\end{bmatrix}, \qquad \mathcal{P}_y=\begin{bmatrix}
1 & 0 & 0\\
0 & -1 & 0\\
0 & 0 & -1
\end{bmatrix}.
\end{equation}
Note, $H_z\to -H_z$ is odd under mirror symmetry since it transforms as a pseudoscalar. One can easily check that parity (mirror) symmetry is broken in both dimensions, $\mathcal{P}_x^{-1} \mathcal{H}(-k_x)\mathcal{P}_x\neq \mathcal{H}(k_x)$ and $\mathcal{P}_y^{-1} \mathcal{H}(-k_y)\mathcal{P}_y\neq \mathcal{H}(k_y)$, when $\Lambda_p\neq 0$. Hence, $\Lambda_p$ transforms exactly as a mass but for spin-1 particles.

Utilizing Maxwell's equations [Eq.~(\ref{eq:MassiveMaxwell})], it is straightforward to derive the dispersion relation of the bulk TM waves,
\begin{equation}
\omega^2-\Lambda_p^2=v_p^2k^2,
\end{equation}
which is identical to the massive Dirac dispersion [Eq.~(\ref{eq:DispersionDirac})]. Rearranging, we obtain the dispersion relation in terms of $\varepsilon$ and $g$ explicitly,
\begin{equation}\label{eq:BulkDispersionRelation}
\omega^2\left(\frac{\varepsilon^2-g^2}{\varepsilon}\right)=\omega^2\varepsilon_\mathrm{eff}=k^2.
\end{equation}
$\varepsilon_\mathrm{eff}$ is the effective permittivity seen by the electromagnetic field,
\begin{equation}
\varepsilon_\mathrm{eff}=\frac{\varepsilon^2-g^2}{\varepsilon}.
\end{equation}
It is clear that whenever $\varepsilon_\mathrm{eff}<0$, electromagnetic waves decay exponentially into the medium. The ``rest energies" are the frequencies at which $\varepsilon_\mathrm{eff}=0$ and define the stationary points $k=0$. This occurs precisely when $\varepsilon^2=g^2$, or equivalently $\omega^2=\Lambda_p^2$.

\subsection{Drude model under an applied magnetic field}

The conventional Drude model, under a biasing magnetic field $B_0$, treats the electron density as an incompressible gas. The Drude model is characterized by two parameters: the plasma frequency $\omega_p$ and the cyclotron frequency $\omega_c=eB_0/M^*$, where $e$ is elementary charge and $M^*$ is the effective mass of the electron. Assuming an applied field in the $-\Hat{z}$ direction, the scalar permittivity $\varepsilon$ and gyrotropic coefficient $g$ are expressed as,
\begin{equation}
\varepsilon=1+\frac{\omega_p^2}{\omega_c^2-\omega^2}, \qquad g=\frac{\omega_c\omega_p^2}{\omega(\omega_c^2-\omega^2)}.
\end{equation}
The effective photonic mass $\Lambda_p$ is therefore,
\begin{equation}
\Lambda_p=\omega\frac{g}{\varepsilon}=\frac{\omega_c\omega_p^2}{\omega_p^2+\omega_c^2-\omega^2}.
\end{equation}
Due to dispersion, the photon sees a different mass at varying frequencies $\omega$ and vanishes at sufficiently high energy $\lim_{\omega\to\infty}\Lambda_p\to 0$. However, the mass is \textit{infinite} $\lim_{\omega\to\omega_0}\Lambda_p\to \infty$ when the frequency is on resonance $\omega_0=\sqrt{\omega_p^2+\omega_c^2}$, which corresponds to the epsilon-near-zero (ENZ) \cite{Andrea2007} condition $\varepsilon(\omega_0)=0$.

The natural eigenmodes of the system $\omega=\omega(k)$, ie. the bulk propagating modes, represent self-consistent solutions to the wave equation, when $k$ and $\omega$ are both real-valued. Plugging our Drude parameters into Eq.~(\ref{eq:BulkDispersionRelation}), we uncover two bulk eigenmode branches $\omega=\omega_\pm$,
\begin{equation}
\omega_\pm^2=\frac{1}{2}\left[2\omega_p^2+\omega_c^2+k^2\pm\sqrt{4\omega_p^2\omega_c^2+(\omega_c^2-k^2)^2}\right].
\end{equation}
$\omega_+$ and $\omega_-$ are the high and low energy eigenmodes respectively. Besides breaking parity and time-reversal, gyrotropy also hybridizes transverse and longitudinal waves. When $\omega_c=0$, the high frequency mode reduces to the transverse ($\vec{k}\cdot\vec{E}=0$) bulk plasmon $\omega_+=\sqrt{\omega_p^2+k^2}$ while the low frequency mode $\omega_-=\omega_p$ reduces to the longitudinal ($\vec{k}\cdot\vec{E}\neq 0$) plasmon. These modes are degenerate at the stationary point $k=0$. However, when $\omega_c\neq 0$, the $\omega_\pm$ bands are fully gapped and the degeneracy at $k=0$ is removed,
\begin{equation}
\omega_\pm(0)=\frac{1}{2}\left|\sqrt{4\omega_p^2+\omega_c^2}\pm\omega_c\right|.
\end{equation}
These represent the rest energies $\varepsilon^2=g^2$ (or $\omega^2=\Lambda_p^2$). Likewise, the asymptotic dependence is,
\begin{equation}
\lim_{k\to\infty}\omega_+\to k, \qquad \lim_{k\to\infty}\omega_-\to \omega_0=\sqrt{\omega_p^2+\omega_c^2}.
\end{equation}
The high energy branch $\omega_+$ approaches the free-photon dispersion where the effective photon mass $\Lambda_p\to0$ vanishes. The low energy branch $\omega_-$ approaches a completely flat dispersion due to an infinite effective mass $\Lambda_p\to\infty$.

\begin{table*}[htbp]
\centering
\caption{Summary of the three unidirectional (chiral) photonic edge states arising in two-dimensional gyroelectric media, with their important properties listed. The quantum gyroelectric effect (QGEE) is a topologically-protected edge state and exists at any boundary - even vacuum. The photonic quantum Hall (PQH) edge state emerges at a perfect magnetic conductor (PMC) boundary condition. These edge states are unique because they carry a high frequency quantum Hall edge current $I_y$. The photonic Jackiw-Rebbi (PJR) edge states are the electromagnetic analogue of the inverted Dirac mass problem and arise at a perfect electric conductor (PEC) boundary condition.}\label{tab:SummEdgeStates}
\begin{tabular}{l|l|l|l|l|l|l|l|l}
\hline
Edge state                         & Boundary condition & Nonlocality & Chiral? & $\mathcal{T}$ broken? & $\mathcal{P}_x$ broken? & $\mathcal{P}_y$ broken? & TEM wave? & Top.-protected? \\ \hline
QGEE & Open: $f(0)=0$ & necessary   & yes     & yes & yes  & yes      & yes ($k\approx 0$)         & yes                      \\ \hline
PQH   & PMC: $\mathcal{P}_x f(-x)=+f(x)$    & unnecessary & yes     & yes & no   & yes               & no  & no                     \\ \hline
PJR        & PEC: $\mathcal{P}_x f(-x)=-f(x)$   & unnecessary & yes     & yes   & no & yes   & yes             & no                       \\ \hline

\end{tabular}
\end{table*}

\section{Quantum gyroelectric effect (QGEE)}\label{sec:QGEE}

\subsection{Topological Drude model}

To make the Drude model \textit{topological} and uncover \textit{topologically-protected} edge states, we need to incorporate spatial dispersion (nonlocality). This purely nonlocal phenomenon has been dubbed the quantum gyroelectric effect (QGEE) and has only been proposed very recently \cite{VanMechelen2018,VanMechelen:19}. A more thorough discussion of temporal and spatial dispersion is provided in Appendix \ref{sec:TemporalDispersion} and \ref{app:Nonlocal}. In the hydrodynamic Drude model, nonlocality emerges when we treat the electron density as a compressible gas. The electron pressure behaves like a restoring force and introduces a first order momentum correction to the longitudinal plasma frequency,
\begin{equation}
(\omega_p^2)_L\to\omega_p^2+\beta^2k^2=(\omega_p+\beta k)^2-2\omega_p\beta k.
\end{equation}
However, topological phases require \textit{second order} momentum corrections at minimum - we must go beyond the hydrodynamic Drude model. Both the plasma frequency,
\begin{equation}
\omega_p\to\Omega_p=\omega_p+\beta_p k^2,
\end{equation}
and the cyclotron frequency, 
\begin{equation}
\omega_c\to\Omega_c=\omega_c+\beta_c k^2,
\end{equation}
must be expanded to second order in $k$. This will alter the behavior of deep subwavelength fields $k\to\infty$ which has very important topological implications. We stress this point as it is imperative to all topological field theories. Spatial dispersion is fundamentally necessary if the electromagnetic theory is to be consistent with the tenfold way \cite{Ryu_2010}, which describes all possible continuum topological phases. A rigorous proof is provided in Appendix \ref{app:ChernNumber}.

Physically, this nonlocal behaviour arises from high momentum corrections to the effective electron mass $M^*$, since the electronic bands are not perfectly parabolic,
\begin{equation}
\frac{1}{M^*}=\frac{1}{\hbar^2}\frac{\partial^2E}{\partial k^2}=\frac{1}{M_0}+\frac{1}{M_2}(ka)^2+\ldots
\end{equation}
$a$ is the lattice constant in this case. The cyclotron frequency corrected to second order $\Omega_c=\omega_c+\beta_ck^2$ is thus,
\begin{equation}
\omega_c=\frac{eB_0}{M_0},\qquad \beta_c=\frac{eB_0 a^2}{M_2}.
\end{equation}
In Appendix~\ref{app:PhotonSpin}, we show that the electromagnetic Chern number $C_\pm$ for each band $\omega=\omega_\pm$, is determined by the relative sign of the cyclotron parameters,
\begin{equation}\label{eq:ChernSecond}
C_\pm=\mp\left[\mathrm{sgn}(\omega_c)-\mathrm{sgn}(\beta_c)\right].
\end{equation}
Alternately, Eq.~(\ref{eq:ChernSecond}) is expressed in terms of the relative signs of the effective electron masses, $M_0$ and $M_2$, and the applied magnetic field $B_0$,
\begin{equation}
C_\pm=\mp[\mathrm{sgn}(M_0)-\mathrm{sgn}(M_2)]\mathrm{sgn}(B_0).
\end{equation}
If $M_0M_2<0$, the electromagnetic phase is topologically nontrivial $|C_\pm|=2$ which requires a change in sign of $1/M^*$ with momentum $k$. In other words, the cyclotron frequency must change sign $\omega_c\beta_c<0$. This implies the electronic band has an \textit{inflection point} at some finite momentum $1/M^*=\partial^2 E/\partial k^2=0$ such that the curvature of the band changes. More precisely, if there are an \textit{odd} number of inflection points, $1/M^*$ changes sign an odd number of times, which always produces $|C_\pm|=2$. It is important to note that a Chern number of $|C|=1$ is only possible when magnetism ($\mu$) is present. All gyrotropic phases possess Chern numbers of $|C|=2$ which is guaranteed by rotational symmetry. A proof is provided in Appendix~\ref{app:PhotonSpin}.

\subsection{Weak magnetic field approximation}

A complete analysis of the topological Drude model warrants its own dedicated paper. Here, we examine only the topological edge states arising in a weak magnetic field $\Omega_c\approx 0$ approximation, at energies far above the cyclotron frequency $\omega\gg \omega_c$. We also ignore any hydrodynamic corrections since they do not affect the topology of the electromagnetic field. The main goal of this section is to demonstrate how nonlocal gyrotropy $g(\omega,k)$ leads to topological phenomena \cite{VanMechelen2018,VanMechelen:19} that can never be realized in a purely local theory.

Assuming $\Omega_c\approx 0$ is sufficiently small and $\omega\gg \omega_c$, we obtain at first approximation ($k\approx 0$),
\begin{equation}
\varepsilon(\omega)\approx 1-\frac{\omega_p^2}{\omega^2}, \qquad g(\omega,k)\approx-\frac{\omega_p^2}{\omega^3}(\omega_c+\beta_c k^2).
\end{equation}
Only the gyrotropic coefficient $g$ adds nonlocal corrections since it is linearly proportional in $\Omega_c$, but is considerably weak. Nevertheless, a unidirectional edge state \textit{always} exists if $\omega_c\beta_c<0$, which corresponds to the topologically nontrivial regime [Eq.~(\ref{eq:ChernSecond})]. We now define,
\begin{equation}
g(\omega,k)= g_0(\omega)-g_2(\omega)k^2,
\end{equation}
with,
\begin{equation}
g_0=-\frac{\omega_c\omega_p^2}{\omega^3}, \qquad g_2=\frac{\beta_c\omega_p^2}{\omega^3}.
\end{equation}
Due to nonlocality in $g$, there are now two characteristic wavelengths $k_{1,2}^2$, which implies two decay channels are active $\eta_{1,2}=\sqrt{k_y^2-k_{1,2}^2}$. The edge state dispersion $\omega=\omega(k_y)$ is determined by the boundary condition which must be insensitive to perturbations at $x=0$. Therefore, we must search for \textit{open boundary solutions} \cite{Medhi_2012,Shen2011,shen2017topological} such that every component of the electromagnetic field vanishes at $x=0$,
\begin{equation}
f(0)=0.
\end{equation}
The open boundary condition is fundamental to topologically-protected edge states. No conventional surface wave, such as SPPs, Dyakonov, Tamm waves, etc. \cite{polo2013electromagnetic} satisfies this constraint since their very existence hinges on the boundary condition. For instance, SPPs intrinsically require a metal-dielectric boundary condition. Conversely, topologically-protected edge states of the QGEE exist at any boundary, since they are defined independent of the contacting medium. This is a statement of bulk-boundary correspondence \cite{Mong2011}.

\subsection{Topologically-protected chiral edge states}

We now impose open boundary conditions on the electromagnetic $f(0)=0$ and look for nontrivial solutions $f(x> 0)\neq 0$ that simultaneously decay into the bulk $f(x\to\infty)\to 0$. Since $f$ contains three components, $E_x$, $E_y$ and $H_z$, the system of equations is overdetermined unless one of the equations can be made linearly dependent on the other two. Based on insight derived from the Dirac equation [Eq.~(\ref{eq:DiracEquation})], we find that the only nontrivial solution requires $E_y(x)=0$. This represents a completely transverse electro-magnetic (TEM) wave as there is no component of the field parallel to the momentum $k_y$. The two decay lengths $\eta_{1,2}$ are roots of the secular equation,
\begin{equation}
\frac{k_y}{\varepsilon}\left(g_0-g_2k_y^2+g_2\eta^2\right)=\eta, \qquad k_y^2=\omega^2\varepsilon,
\end{equation}
which produces,
\begin{equation}\label{eq:Roots}
\eta_{1,2}=\frac{1}{2g_2}\left[\frac{\varepsilon}{k_y}\pm \sqrt{\left(\frac{\varepsilon}{k_y}\right)^2+4g_2(g_2k_y^2-g_0)}\right].
\end{equation}
Notice that an edge state only exists when $\varepsilon>0$ is positive. This is very different from SPPs which require a negative permittivity. For our weak field approximation, the edge dispersion is simply,
\begin{equation}
\omega^2=\omega_p^2+k_y^2.
\end{equation}
A solution always exists whenever $k_y^2<g_0g_1>0$ such that both $\Re[\eta_{1,2}]>0$ are decaying modes. This criteria is only satisfied in the topologically nontrivial regime $\omega_c\beta_c<0$, confirming our theory. $\mathrm{sgn}(\omega_c)=\mathrm{sgn}(-\beta_c)=+1$ is a backward propagating wave while $\mathrm{sgn}(\omega_c)=\mathrm{sgn}(-\beta_c)=-1$ is forward propagating. The edge state is completely unidirectional (chiral) since $k_y\to -k_y$ cannot be a simultaneous solution. Back-scattering is forbidden.

After a bit of work, we obtain the final expression for the (low momentum) topologically-protected edge state,
\begin{equation}
f(x\geq 0)=\begin{bmatrix}
E_x\\E_y\\H_z
\end{bmatrix}=f_0\Big(\hat{x}-s_{k_y}\sqrt{\varepsilon}\hat{z} \Big)\left(e^{ -\eta_1 x}-e^{ -\eta_2 x}\right).
\end{equation}
$s_{k_y}=\mathrm{sgn}(k_y)$ is the sign of the momentum which dictates the direction of propagation and $f_0$ is a proportionality constant. Remarkably, the edge wave behaves identically to a vacuum photon (completely transverse polarized) but with a modified dispersion. Indeed, they are \textit{helically quantized} along the direction of propagation $\hat{k}=\Hat{y}$. This is the definition of longitudinal spin-momentum locking as $f$ is an \textit{eigenstate} of $\Hat{S}_y$,
\begin{equation}\label{eq:QGEEedgestate}
\Hat{S}_y\mathfrak{F}=
s_{k_y}\mathfrak{F},\qquad\mathfrak{F}=\begin{bmatrix}
\sqrt{\varepsilon}E_x\\
\sqrt{\varepsilon}E_y\\
H_z
\end{bmatrix}.
\end{equation}
$\Hat{k}\cdot\vec{S}=\Hat{S}_y$ is the helicity operator along $\Hat{y}$, which was defined in Eq.~(\ref{eq:Spin1Operators}). Notice that the spin is quantized $s_{k_y}=\mathrm{sgn}(k_y)=\pm 1$ and completely locked to the momentum as it depends on the direction of propagation. A summary of the QGEE and its important properties is listed in Tab.~\ref{tab:SummEdgeStates}.

\begin{figure*}
\includegraphics[width=\linewidth]{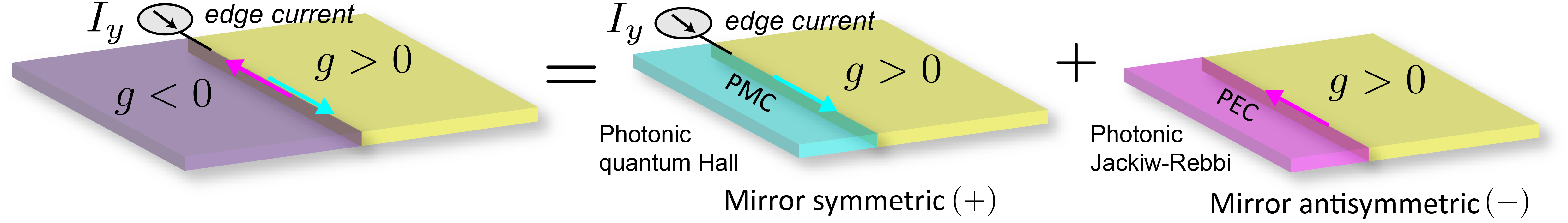}
\caption{The interface of two optical isomers with positive $+g$ and negative $-g$ gyrotropy. In the Drude model, this corresponds to reversed magnetic biasing $\pm B_0$. The interface hosts two edge states that can be decomposed into two chiral (unidirectional) subsystems with perfect magnetic conductor (PMC) and perfect electric conductor (PEC) boundary conditions. PMC and PEC are mirror symmetric $(+)$ and mirror antiysmmetric $(-)$ respectively,
designating photonic quantum Hall (PQH) and photonic Jackiw-Rebbi (PJR) states. The particular mirror symmetry $(\pm)$ dictates how the electromagnetic field transforms into the virtual photon $\mathcal{P}_xf (-x)=\pm f (x)$.}
\label{fig:MirrorBC}
\end{figure*}

\section{Interface of optical isomers}\label{sec:OpticalIsomers}

In Sec.~\ref{sec:QGEE}, we showed that nonlocal gyrotropy $g(\omega,k)$ can lead to topologically-protected chiral edge states that satisfy open boundary conditions. In the Drude model, this arises from a momentum dependent cyclotron frequency $\Omega_c(k)=\omega_c+\beta_ck^2$ that changes sign within the dispersion $\omega_c\beta_c<0$. Discovering such a material and observing these topological edge waves remains an open problem. Here, we consider a more practical scenario that does not involve nonlocality $\beta_c=0$, but hosts intriguing physics nonetheless.

Instead of having $g$ change sign with momentum, we let $g$ vary with position $g\to g(x)$ such that it defines the boundary between two distinct materials. The simplest case represents the boundary of two ``optical isomers" \cite{Zhukov2000,Zyuzin2015}, with $g$ in the $x>0$ space and $-g$ in the $x<0$ space but $\varepsilon$ identical in both media. The permittivity tensors are therefore complex conjugates of one another $\varepsilon_{ij}(x)=\varepsilon_{ij}^*(-x)$ and there is perfect mirror symmetry about $x=0$. In the Drude model, this represents the interface between two biased plasmas, but with reversed applied fields $\pm B_0$. The cyclotron frequencies in each half-space are exactly opposite $\pm \omega_c=\pm eB_0/M_0$. Note though, this implies the biasing field is discontinuous across the boundary $B_0(0^+)\neq B_0(0^-)$ which is an idealization. In reality, there must be a field gradient $B_0\to B_0(x)$ that interpolates between the two regions. However, we get this desired behavior for free if we assume a perfect mirror in the $x<0$ half-space, such that the virtual photon is the exact mirror image \cite{Horsley2018}. This is because the permittivity is even under mirror symmetry $\varepsilon\to\varepsilon$ while gyrotropy is odd $g\to -g$.

There are two types of mirrors we can introduce: a perfect magnetic conductor (PMC) or a perfect electric conductor (PEC). The difference between the two lies in the type of symmetry of the boundary condition. PMC represents symmetric $(+)$ boundary conditions and PEC is antisymmetric $(-)$. Under each symmetry ($\pm$) the electromagnetic field $f$ must transform into its mirror image as $\mathcal{P}_xf(-x)=\pm f(x)$. As we will see, each mirror has a chiral (unidirectional) edge state associated with it, but with very different properties. A visualization of the two mirror boundary conditions is displayed in Fig.~\ref{fig:MirrorBC}. It must be stressed that a \textit{real} interface of optical isomers hosts both edge states. A symmetric (PMC) state propagates in one direction while the antisymmetric (PEC) state propagates in the opposite direction. Only when we enforce a specific boundary condition can we isolate for either edge state.

\section{Photonic quantum Hall (PQH) edge states}\label{sec:QuantumHall}

The photonic quantum Hall (PQH) edge states are symmetric (PMC) solutions of the optical isomer problem. These states are unique in that they support a high frequency quantum Hall edge current at the interface. The first step is to derive the $\delta$-potential characterizing the potential energy at the discontinuity $x=0$. This arises from a sudden change in the gyrotropic coefficient $g\to g~\mathrm{sgn}(x)$. Assuming the longitudinal field is nonzero $E_y\neq 0$, it can be shown that $E_y$ satisfies a Schr\"{o}dinger-like wave equation,
\begin{equation}\label{eq:LocalDelta}
-\partial_x^2E_y+V(x)E_y=\mathcal{E}E_y.
\end{equation}
$V(x)$ is the ``potential energy" and after differentiating reduces to a $\delta$-function,
\begin{equation}
V(x)=k_y\frac{g}{\varepsilon}\partial_x\mathrm{sgn}(x)=2k_y\frac{g}{\varepsilon}\delta(x).
\end{equation}
$\mathcal{E}$ is the corresponding ``energy eigenvalue",
\begin{equation}
 \mathcal{E}=\omega^2\left(\varepsilon-\frac{g^2}{\varepsilon}\right)-k_y^2.
\end{equation}
It is well known that $\delta$-potentials always possess a bound state when the potential energy is attractive $V(x)<0$. Therefore, $k_yg/\varepsilon<0$ must always be satisfied for any given frequency and wave vector. The chirality of the bound state is immediately apparent. If a solution exists for a particular $k_y$, then $k_y\to-k_y$ is never a simultaneous solution. Back-scattering is forbidden.

To solve Eq.~(\ref{eq:LocalDelta}), we integrate both sides of the equation from $\int_{0^-}^{0^+}dx$ while assuming $E_y(x)=E_y(0)\exp(-\eta|x|)$. In this case, the longitudinal electric field is continuous across the domain wall $E_y(0^+)=E_y(0^-)$. We obtain a surprisingly simple characteristic equation,
\begin{equation}\label{eq:LocalCharacteristicEquation}
\eta=-k_y \frac{g}{\varepsilon},\qquad k_y^2=\omega^2\varepsilon.
\end{equation}
Notice that an edge state only exists when $\varepsilon>0$ is positive. This is very different from SPPs which require a negative permittivity. After some algebra, the $E_x$ and $H_z$ fields can be expressed as,
\begin{subequations}
\begin{equation}
E_x(x)=-is_x \frac{\varepsilon^2+g^2}{2\varepsilon g}E_y(0)e^{-\eta|x|},
\end{equation}
\begin{equation}
H_z(x)=is_xs_{k_y} \frac{\varepsilon^2-g^2}{2\sqrt{\varepsilon}g}E_y(0)e^{-\eta|x|},
\end{equation}
\end{subequations}
where $s_x=\mathrm{sgn}(x)$ and $s_{k_y}=\mathrm{sgn}(k_y)$ denotes the sign of $x$ and $k_y$ respectively. It is easy to check that the PQH state is mirror symmetric $\mathcal{P}_xf(-x)=+f(x)$ about $x=0$.

However, one might expect the normal electric field $E_x$ and tangential magnetic field $H_z$ to vanish at $x=0$ due to PMC boundary conditions. This is not the case. A free edge current is running parallel to the interface, such that the fields are discontinuous,
\begin{equation}
I_y=\frac{1}{2}\left[H_z(0^-)-H_z(0^+)\right]=-is_{k_y} \frac{\varepsilon^2-g^2}{2\sqrt{\varepsilon}g}E_y(0).
\end{equation}
Note, we divide by a factor of 2 to remove the contribution from the virtual photon. $I_y$ is the high frequency analogue of the quantum Hall edge current. Interestingly, these photonic edge waves can be excited by passing a time-varying current along the boundary - similar to a transmission line \cite{cheng2013field}. However, current can only flow in one direction and the system behaves like a simultaneous photonic and electronic diode.

\begin{figure}
\includegraphics[width=\linewidth]{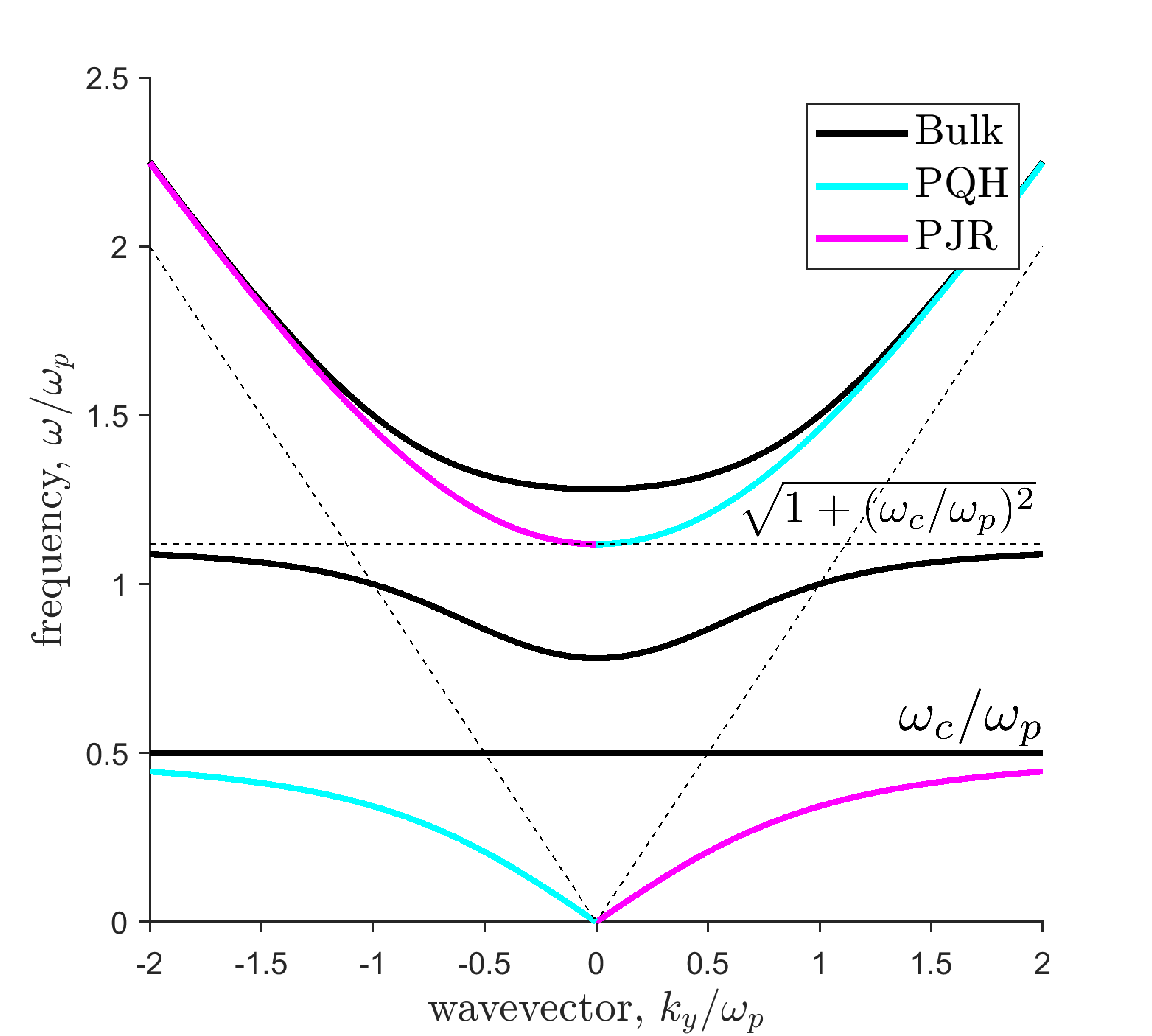}
\caption{Dispersion relation of the local Drude model under an applied magnetic field with $\omega_c/\omega_p=1/2$ as an example. Black lines indicate bulk bands while cyan and magnetic lines represent unidirectional photonic quantum Hall (PQH) and photonic Jackiw-Rebbi (PJR) edge states respectively. There are a total of 3 positive energy bulk bands. Two correspond to high and low frequency TM modes $\omega=\omega_\pm$ while the third represents pure cyclotron orbits $\omega=\omega_c$. The PQH states emerge at a PMC boundary while the PJR states require a PEC boundary. Unlike conventional SPPs, the PQH and PJR states asymptotically approach the bulk bands in the $k_y\to \infty$ limit. The upper branch approaches the free photon dispersion $\omega_\uparrow\to k_y$ while the lower branch approaches pure cyclotron orbits $\omega_\downarrow\to \omega_c$. The frequency range where \textit{no} edge state exists $\omega_c<\omega<\sqrt{\omega_p^2+\omega_c^2}$, corresponds to the plasmonic region $\varepsilon<0$. Indeed, these edge waves are fundamentally different from SPPs as they require $\varepsilon>0$.}
\label{fig:LocalDispersion}
\end{figure}

Now we look for self-consistent solutions to the dispersion relation [Eq.~(\ref{eq:LocalCharacteristicEquation})] which correspond to propagating edge modes, with both $k_y$ and $\omega$ real-valued. There are in fact two edge bands which span the gaps between the bulk bands,
\begin{equation}
\omega^2_{\uparrow\downarrow}=\frac{1}{2}\left[\omega_p^2+\omega_c^2+k_y^2\pm \sqrt{(\omega_p^2+\omega_c^2+k_y^2)^2-4k_y^2\omega_c^2}\right].
\end{equation}
$\omega_\uparrow$ spans the region between the upper $\omega_+$ and lower $\omega_-$ bulk TM bands while $\omega_\downarrow$ spans between $\omega_c$ and $0$. Now we need to check when $\eta>0$ represents a decaying wave for the two edge modes,
\begin{equation}
\eta_{\uparrow\downarrow}=-k_y\frac{g(\omega_{\uparrow\downarrow})}{\varepsilon(\omega_{\uparrow\downarrow})}=\frac{k_y\omega_c\omega_p^2}{\omega_{\uparrow\downarrow}(\omega_{\uparrow\downarrow}^2-\omega_p^2-\omega_c^2)}.
\end{equation}
Since $\omega_\uparrow^2\geq \omega_p^2+\omega_c^2$ for all $k_y$, then $\omega_c k_y>0$ must always be satisfied in the $\omega_\uparrow$ frequency region. Choosing $\omega_c>0$, the upper edge branch propagates strictly in the $k_y>0$ direction. Similarly, since $\omega_\downarrow^2< \omega_p^2+\omega_c^2$ for all $k_y$, then $\omega_c k_y<0$ must always be satisfied in the $\omega_\downarrow$ frequency region. The lower edge branch propagates strictly in the $k_y<0$ direction. The dispersion relation of the PQH edge states are displayed in Fig.~\ref{fig:LocalDispersion}.

\section{Photonic Jackiw-Rebbi (PJR) edge states}\label{sec:JackiwRebbi}

The photonic Jackiw-Rebbi (PJR) edge states are antisymmetric (PEC) solutions of the optical isomer problem. Like the QGEE, these edge states are completely transverse electro-magnetic (TEM) waves. PJR states share many important properties with the QGEE [Sec.~\ref{sec:QGEE}] even though they arise by a very different means. The only significant difference is that they do not satisfy open boundary conditions and necessarily require a PEC boundary. This means they are not topologically-protected as they are sensitive to boundary conditions. However, this particular system is the most practical experimentally.

To solve, we first assume the magnetic field is continuous across the domain wall $H_z(0^+)=H_z(0^-)$ such that zero edge current $I_y=0$ is excited. We obtain an identical dispersion relation as the PQH states [Eq.~(\ref{eq:LocalCharacteristicEquation})], except the wave propagates in the reverse direction,
\begin{equation}\label{eq:Characteristic_Equation}
\eta=k_y\frac{g}{\varepsilon}, \qquad k_y^2=\omega^2\varepsilon.
\end{equation}
There is an immediate connection with the Dirac Jackiw-Rebbi dispersion [Eq.~(\ref{eq:DiracJRDispersion})], with respect to the effective speed of light $v_p$ and effective photon mass $\Lambda_p$,
\begin{equation}
\eta=\frac{\omega g}{\sqrt{\varepsilon}}=\frac{|\Lambda_p|}{v_p}, \qquad \omega^2=\frac{k_y^2}{\varepsilon}=v_p^2k_y^2.
\end{equation}
Surprisingly, the electromagnetic field profile of the PJR state is drastically different than the PQH state. The longitudinal field vanishes $E_y(x)=0$ entirely because $E_y(0^+)=E_y(0^-)=0$ is required by symmetry. Hence, the PEC states correspond to completely transverse electro-magnetic (TEM) edge waves,
\begin{equation}
f(x)=E_x(0)\Big(\hat{x}-s_{k_y}\sqrt{\varepsilon}\hat{z} \Big)e^{ -\eta |x|}.
\end{equation}
It is easy to check that the PJR state is mirror antisymmetric $\mathcal{P}_xf(-x)=-f(x)$ about $x=0$. The edge wave behaves identically to a vacuum photon (transverse polarized) but with a modified dispersion. Indeed, they are \textit{helically quantized} along the direction of propagation $\Hat{k}\cdot \Vec{E}=\Hat{y}\cdot \Vec{E}=E_y=0$. This is the definition of longitudinal spin-momentum locking as $f$ is an \textit{eigenstate} of $\Hat{S}_y$,
\begin{equation}\label{eq:SMLPhoton}
\Hat{S}_y\mathfrak{F}=
s_{k_y}\mathfrak{F},\qquad\mathfrak{F}=\begin{bmatrix}
\sqrt{\varepsilon}E_x\\
\sqrt{\varepsilon}E_y\\
H_z
\end{bmatrix}.
\end{equation}
$\Hat{k}\cdot\vec{S}=\Hat{S}_y$ is the helicity operator along $\Hat{y}$, which was defined in Eq.~(\ref{eq:Spin1Operators}). Notice that the spin is quantized $s_{k_y}=\mathrm{sgn}(k_y)=\pm 1$ and completely locked to the momentum as it depends on the direction of propagation. This should be contrasted with their electron (spin-\sfrac{1}{2}) equivalent in Eq.~(\ref{eq:SMLDirac}). The dispersion relation of the PJR edge states are displayed in Fig.~\ref{fig:LocalDispersion}. A short discussion on the robustness of PQH and PJR states is presented in Appendix \ref{app:Robust}.

\section{Conclusion}\label{sec:Conclusions}

In summary, we have identified the three fundamental classes of unidirectional photonic edge waves arising in gyroelectric media. The quantum gyroelectric effect (QGEE) is a topologically-protected edge state that requires nonlocal gyrotropy. This wave satisfies open boundary conditions and displays bulk-boundary correspondence as it is defined independent of the contacting medium. The photonic quantum Hall (PQH) and photonic Jackiw-Rebbi (PJR) states are local phenomena and emerge at the interface of optical isomers - two media with inverted gyrotropy.


\section*{Acknowledgements}

This research was supported by the Defense Advanced Research Projects Agency (DARPA) Nascent Light-Matter Interactions (NLM) Program and the National Science Foundation (NSF) [Grant No. EFMA-1641101].

\appendix

\section*{Appendix}

\section{Dirac Jackiw-Rebbi edge states}\label{app:JRDirac}

For completeness, we provide a brief review of Jackiw-Rebbi states that arise in two-dimensional condensed matter systems. The simplest realization is described by the 2D Dirac equation $H\Psi=E\Psi$,
\begin{equation}\label{eq:DiracEquation}
H=v(k_x\sigma_x+k_y\sigma_y)+\Lambda\sigma_z,
\end{equation}
where $[\sigma_i,\sigma_j]=2i\epsilon_{ijk}\sigma_k$ are the Pauli matrices. $v$ is the Fermi velocity and $\Lambda$ is a two-dimensional Dirac mass.

We consider an interface of two Dirac particles with opposite masses $\Lambda\to \Lambda\mathrm{sgn}(x)$. Similar to the photonic problem [Sec.~\ref{sec:OpticalIsomers}], there is now mirror symmetry about $x=0$. The unidirectional (chiral) edge solution is well known \cite{Shen2011} and assumes a surprisingly simple form,
\begin{equation}
\Psi(x)=\Psi_0\begin{bmatrix}
1\\ i~ s_{k_y}
\end{bmatrix}e^{-\eta|x|},
\end{equation}
where $s_{k_y}=\mathrm{sgn}(k_y)=\pm 1$ is the sign of the momentum. This follows from the characteristic equation,
\begin{equation}\label{eq:DiracJRDispersion}
\eta=\frac{|\Lambda|}{v}, \qquad E^2=v^2k_y^2.
\end{equation}
If $\Lambda>0$, the Dirac edge wave propagates strictly in the $k_y>0$ direction and vice verse for $\Lambda<0$. It is clear that $\Psi$ is an eigenstate of both the helicity operator $\Hat{k}\cdot\vec{S}=\sigma_y/2$ and the mirror operator $\mathcal{P}_x=\sigma_y$ which are identical in this case,
\begin{equation}\label{eq:SMLDirac}
\sigma_y\Psi(-x)=\sigma_y\Psi(x)=s_{k_y}\Psi(x).
\end{equation}
Indeed, the Dirac Jackiw-Rebbi edge states are helically quantized and behave identically to a massless (Weyl) fermion. This should be contrasted with their photonic (spin-1) equivalent in Eq.~(\ref{eq:SMLPhoton}).

\section{Robustness of PQH and PJR edge states}\label{app:Robust}

Although the PQH and PJR states are not topologically-protected, they can still exhibit robust transport - ie. immunity to small perturbations in the gyrotropic coefficient $g$. Let us assume $g\to g(x)$ is a function of $x$ but take $\varepsilon$ as a constant in space. In reality, this is only approximately true since $g$ and $\varepsilon$ cannot be completely independent functions. In the Drude model for instance, a field gradient $B_0\to B_0(x)$ creates a spatially dependent cyclotron frequency $\omega_c\to\omega_c(x)$ which alters both the resonance frequency and the relative magnitude of the gyrotropy. Hence, both $g$ and $\varepsilon$ will generally vary with $x$. However, this simplifying assumption illustrates the point very well and holds for relatively small perturbations in the gyrotropy.

When only $g(x)$ varies with $x$, the Schr\"{o}dinger-like wave equation [Eq.~(\ref{eq:LocalDelta})] for the PQH state becomes,
\begin{equation}\label{eq:SpatialPotential}
-\partial_x^2E_y+\left[\frac{k_y}{\varepsilon}\partial_xg(x)+\omega^2\frac{g^2(x)}{\varepsilon} \right]E_y=(\omega^2\varepsilon-k_y^2)E_y.
\end{equation}
Due to the mirror boundary condition, $g(-x)=-g(x)$ is an odd function of $x$. However, we can still allow a jump discontinuity at $x=0$, such that $g(0^-)=-g(0^+)$. Far from the boundary $|x|\to\infty$, the gyrotropy approaches the uniform bulk $g(x\to\pm\infty)=\pm g_0$. A unidirectional edge state always exists and is immune to perturbations in $g$. To prove this, we choose an integrating factor of the form,
\begin{equation}
E_y(x)=E_y(0)\exp\left[\frac{k_y}{\varepsilon}\int_{-\infty}^xg(x')dx'\right],
\end{equation}
which satisfies,
\begin{equation}
\partial_xE_y(x)=\frac{k_y}{\varepsilon}g(x)E_y(x),
\end{equation}
and,
\begin{equation}
\partial_x^2E_y(x)=\left[\frac{k_y}{\varepsilon}\partial_x g(x)+\frac{k_y^2}{\varepsilon^2}g^2(x)\right]E_y(x).
\end{equation}
Clearly, if the edge dispersion is fulfilled $k_y^2=\omega^2\varepsilon$, Eq.~(\ref{eq:SpatialPotential}) is satisfied regardless of the particular form of $g(x)$. The exact same integrating solution exists for the PJR states, with $E_y(x)=0$, except the momentum is reversed $k_y\to-k_y$.

As an example, let $g(x)=g_0\tanh(x/a)$, where $a$ is some characteristic transition length that interpolates between $g(0)=0$ and $g(x\to \pm\infty)=\pm g_0$. The integral of which is $\int g(x')dx'=ag_0\log[\cosh(x/a)]$. The spatial profile then becomes,
\begin{equation}
E_y(x)=E_y(0)[\cosh(x/a)]^{(k_yag_0/\varepsilon)}.
\end{equation}
In the limit of an infinitesimally narrow transition width $a\to 0$, the solution reduces to the idealized case $[\cosh(x/a)]^{(k_yag_0/\varepsilon)}\to \exp(-\eta |x|)$ with $\eta=-k_yg_0/\varepsilon$.

\section{Temporal dispersion}\label{sec:TemporalDispersion}

Temporal dispersion, or the frequency dependence of linear response, arises whenever light couples to matter,
\begin{equation}
\mathcal{M}(\omega)=\begin{bmatrix}
\varepsilon_{xx} & \varepsilon_{xy} & \chi_{x} \\
\varepsilon_{xy}^* & \varepsilon_{yy} & \chi_{y}\\
\chi_{x}^*& \chi_{y}^* & \mu
\end{bmatrix}, ~~ \begin{array}{ll}
D_i=\varepsilon_{ij}E^j+\chi_{i} H_z, \\
~\\
B_z=\chi_{i}^*E^i+\mu H_z.\\
\end{array}
\end{equation}
Temporal dispersion is always present because it characterizes the relative coupling at a particular energy to the material degrees of freedom - the electronic modes. These are the physical objects that generate the linear response theory to begin with. Moreover, due to the reality condition of the electromagnetic field (particle-antiparticle symmetry), the real and imaginary components of $\mathcal{M}$ cannot be arbitrary functions of $\omega$,
\begin{equation}
\mathcal{M}^*(-\omega)=\mathcal{M}(\omega).
\end{equation}
This implies $\Re[\mathcal{M}(-\omega)]=\Re[\mathcal{M}(\omega)]$ must be even in $\omega$ while $\Im[\mathcal{M}(-\omega)]=-\Im[\mathcal{M}(\omega)]$ is odd. Hence, it is physically \textit{impossible} to break time-reversal ($\mathcal{T}$) symmetry without dispersion \cite{note1}. In this case, we imply breaking $\mathcal{T}$ symmetry nontrivially (Hermitian response). Adding loss (antiHermitian response) breaks $\mathcal{T}$ symmetry in a trivial way because it does not alter the dynamics of the field - it simply adds a finite lifetime.

Besides the reality condition, $\mathcal{M}$ must satisfy three additional physical constraints. The first being transparency at high frequency,
\begin{equation}
\lim_{\omega\to\infty}\mathcal{M}(\omega)\to\mathds{1}_3,
\end{equation}
where $\mathds{1}_3$ is the $3\times 3$ identity. The second being Kramers-Kronig (causality),
\begin{equation}
\oint_{\Im[\omega']\geq 0} \frac{\mathcal{M}(\omega')-\mathds{1}_3}{\omega'-\omega}d\omega'=0.
\end{equation}
This ensures the response function is analytic in the upper complex plane and decays at least as fast as $|\omega|^{-1}$. The last condition requires a positive definite energy density,
\begin{equation}\label{eq:EnergyDensity}
\bar{\mathcal{M}}(\omega)=\partial_\omega [\omega\mathcal{M}(\omega)]>0.
\end{equation}
By combining all the aforementioned constraints and assuming Hermitian (lossless) systems $\mathcal{M}^\dagger=\mathcal{M}$, we can always expand $\mathcal{M}$ via a partial fraction decomposition,
\begin{equation}\label{eq:OscillatorExp}
\mathcal{M}(\omega)=\mathds{1}_3-\sum_\alpha\frac{\mathcal{C}_\alpha^\dagger \mathcal{C}_\alpha}{\omega_\alpha(\omega-\omega_\alpha)}.
\end{equation}
The poles of the response function $\omega=\omega_\alpha$ represent resonances of the material degrees of freedom. From an electronic band structure point of view, $\omega_\alpha=(E_\alpha-E_0)/\hbar$ represents the energy difference between the ground state and an excited state. $\mathcal{C}_\alpha$ is the coupling strength (matrix element) of the excitation.

\section{Spatial dispersion (nonlocality)}\label{app:Nonlocal}

Spatial dispersion, or the momentum dependence of linear response, dictates how the light-matter interaction changes with wavelength (scale). Nonlocality becomes relevant at the nanoscale and governs the deep subwavelength physics. Perhaps more importantly, nonlocality is fundamentally necessary to describe topological phenomena. As has been proven in Ref.~\cite{VanMechelen2018,VanMechelen:19}, Chern numbers are only quantized when $\mathcal{M}$ is \textit{regularized} which inherently requires spatial dispersion. This is the only way for the electromagnetic theory to be consistent with the tenfold way \cite{Ryu_2010}, which describes all possible continuum topological theories. Technically, the photon belongs to Class D, the same universality class as the $p+ip$-wave topological superconductor \cite{Read2000}. Class D possesses an integer topological invariant (Chern number) in two dimensions.

Spatial dispersion is easily introduced by letting $\omega_{\alpha}\to\omega_{\alpha\mathbf{k}}$ and $\mathcal{C}_{\alpha}\to\mathcal{C}_{\alpha\mathbf{k}}$ be functions of $\mathbf{k}$,
\begin{equation}\label{eq:Nonlocal}
\mathcal{M}(\omega,\mathbf{k})=\mathds{1}_3-\sum_\alpha\frac{\mathcal{C}_{\alpha\mathbf{k}}^\dagger \mathcal{C}_{\alpha\mathbf{k}}}{\omega_{\alpha\mathbf{k}}(\omega-\omega_{\alpha\mathbf{k}})}.
\end{equation}
The $\mathbf{k}$ dependence cannot be completely arbitrary because the response function must satisfy the generalized reality condition,
\begin{equation}
\mathcal{M}^*(-\omega,-\mathbf{k})=\mathcal{M}(\omega,\mathbf{k}).
\end{equation}
The reality condition (particle-antiparticle symmetry) implies there is a negative energy resonance $-\omega_{\alpha-\mathbf{k}}$ associated with each positive energy $\omega_{\alpha\mathbf{k}}$. The wave equation of the 2D photon coupled to matter is thus,
\begin{equation}\label{eq:WaveEquation}
\mathcal{H}_0(\mathbf{k})f=\omega\mathcal{M}(\omega,\mathbf{k})f.
\end{equation}
However, this is still not a first-order eigenvalue problem since $\mathcal{M}$ depends on the eigenvalue $\omega$ itself. Moreover, the electromagnetic field $f$ is not the complete eigenvector of this system. A simple reason is because the number of eigenmodes $n$ should match the dimensionality of the eigenvector $\dim[u]=n$. This clearly does not hold $\dim [f]=3$ when temporal dispersion is present since there can be many modes that satisfy the wave equation [Eq.~(\ref{eq:WaveEquation})].

\subsection{Electromagnetic Hamiltonian}

To convert Eq.~(\ref{eq:WaveEquation}) into a first-order Hamiltonian, we define the auxiliary variables $\psi_\alpha$ that describe the internal polarization and magnetization modes of the medium,
\begin{equation}\label{eq:oscillator_coupling}
\psi_\alpha=\frac{\mathcal{C}_{\alpha\mathbf{k}}f}{\omega-\omega_{\alpha\mathbf{k}}}, \qquad \omega \psi_\alpha=\omega_{\alpha\mathbf{k}}\psi_\alpha+\mathcal{C}_{\alpha\mathbf{k}}f.
\end{equation}
Back-substituting into Eq.~(\ref{eq:WaveEquation}) and using the partial fraction expansion,
\begin{equation}
\frac{\omega}{\omega_\alpha(\omega-\omega_\alpha)}=\frac{1}{\omega_\alpha}+\frac{1}{\omega-\omega_\alpha},
\end{equation}
we obtain the first-order wave equation,
\begin{equation}
H(\mathbf{k})u=\omega u, \qquad u=\begin{bmatrix}
f \\ \psi_1 \\ \psi_2 \\ \vdots
\end{bmatrix}.
\end{equation}
$u$ accounts for the electromagnetic field $f$ and all internal polarization modes $\psi_\alpha$ describing the linear response. $H(\mathbf{k})$ is the Hamiltonian matrix that acts on this generalized state vector $u$,
\begin{equation}
H(\mathbf{k})=\begin{bmatrix}
\mathcal{H}_0(\mathbf{k})+\sum_\alpha \omega_{\alpha\mathbf{k}} ^{-1}\mathcal{C}^\dagger_{\alpha\mathbf{k}} \mathcal{C}_{\alpha\mathbf{k}} & ~~\mathcal{C}^\dagger_{1\mathbf{k}}~~ & ~~\mathcal{C}^\dagger_{2\mathbf{k}}~~ & \ldots \\
\mathcal{C}_{1\mathbf{k}} & \omega_{1\mathbf{k}} & 0 & \ldots\\
\mathcal{C}_{2\mathbf{k}} & 0 & \omega_{2\mathbf{k}} & \ldots\\
\vdots & \vdots & \vdots & \ddots 
\end{bmatrix}.
\end{equation}
This decomposition makes intuitive sense. The dimensionality of the Hamiltonian matches the number of distinct eigenmodes and eigenenergies of the problem. The complete set of eigenvectors is thus,
\begin{equation}
H(\mathbf{k})u_{n\mathbf{k}}=\omega_{n\mathbf{k}}u_{n\mathbf{k}}.
\end{equation}
Constructing the total Hamiltonian $H(\mathbf{k})$ is a very important procedure when nonlocality is present. This is because we have to start imposing boundary conditions on the oscillators $\psi_\alpha$ themselves when we consider interface effects.

Utilizing the linear response theory, the electromagnetic eigenstates of the medium $f_{n\mathbf{k}}$ are solutions of the self-consistent wave equation,
\begin{equation}
\det[\omega\mathcal{M}(\omega,\mathbf{k})-\mathcal{H}_0(\mathbf{k})]=0, \qquad \omega=\omega_{n\mathbf{k}},
\end{equation}
which determines all possible polaritonic bands. These bands are normalized to the energy density as,
\begin{equation}
u^\dagger_{n\mathbf{k}}u_{n\mathbf{k}}=f^\dagger_{n\mathbf{k}}\bar{\mathcal{M}}(\omega_{n\mathbf{k}},\mathbf{k})f_{n\mathbf{k}}=1, 
\end{equation}
where,
\begin{equation}
\bar{\mathcal{M}}(\omega,\mathbf{k})=\partial_\omega[\omega\mathcal{M}(\omega,\mathbf{k})]=\mathds{1}_3+\sum_\alpha\frac{\mathcal{C}^\dagger_{\alpha\mathbf{k}}\mathcal{C}_{\alpha\mathbf{k}} }{(\omega-\omega_{\alpha\mathbf{k}})^2}.
\end{equation}
Due to the constraints on $\mathcal{M}$, these bands are continuous and real-valued for all $\mathbf{k}$. 

\subsection{Nonlocal regularization}

A well known requirement of any continuum topological theory, is that the Hamiltonian must approach a \textit{directionally independent} value in the asymptotic limit \cite{Ryu_2010},
\begin{equation}
\lim_{k\to\infty} H(\mathbf{k})=H(k).
\end{equation}
This ensures the Hamiltonian is connected at infinity and is the continuum equivalent of a periodic boundary condition. Mathematically, this means the momentum space manifold is compact and can be projected onto the Riemann sphere $\mathbb{R}^2\to S^2$. Alternatively, if the response function is regularized, the wave equation approaches a directionally independent value in the asymptotic limit,
\begin{equation}
\lim_{k\to\infty}[\omega\mathcal{M}(\omega,\mathbf{k})-\mathcal{H}_0(\mathbf{k})]\to\omega\mathcal{M}(\omega,k).
\end{equation}
This places constraints on the asymptotic behavior of the response parameters,
\begin{equation}
\lim_{k\to\infty}\mathcal{C}_{\alpha\mathbf{k}}\to \mathcal{C}_{\alpha p} k^p, \qquad \lim_{k\to\infty}\omega_{\alpha\mathbf{k}}\to\omega_{\alpha p}k^p.
\end{equation}
Consequently, $\mathcal{C}_{\alpha\mathbf{k}}$ and $\omega_{\alpha\mathbf{k}}$ require quadratic order nonlocality \textit{at minimum} $p \geq 2$ to be properly regularized. We will show that this is a necessary and sufficient condition for Chern number quantization. 

\section{Continuum electromagnetic Chern number}\label{app:ChernNumber}

The Berry connection is found by varying the complete eigenvectors $u_{n\mathbf{k}}$ with respect to the momentum $\mathbf{A}_n(\mathbf{k})=-iu^\dagger_{n\mathbf{k}}\partial_\mathbf{k}u_{n\mathbf{k}}$. This can be simplified to,
\begin{equation}
\mathbf{A}_n(\mathbf{k})=-if^\dagger_{n\mathbf{k}}\bar{\mathcal{M}}(\omega_{n\mathbf{k}},\mathbf{k})\partial_\mathbf{k} f_{n\mathbf{k}}+f^\dagger_{n\mathbf{k}}\pmb{\mathcal{A}}(\omega_{n\mathbf{k}},\mathbf{k}) f_{n\mathbf{k}},
\end{equation}
where $\pmb{\mathcal{A}}$ is the Berry connection arising from the material degrees of freedom,
\begin{equation}
\pmb{\mathcal{A}}(\omega,\mathbf{k})=-i\sum_\alpha \frac{\mathcal{C}^\dagger_{\alpha\mathbf{k}}\partial_\mathbf{k}\mathcal{C}_{\alpha\mathbf{k}}}{(\omega-\omega_{\alpha\mathbf{k}})^2}.
\end{equation}
It is straightforward to prove that nonlocal regularization guarantees Chern number quantization. In the asymptotic limit, the electromagnetic modes approach a directionally independent value, up to a possible U(1) gauge,
\begin{equation}
\lim_{k\to\infty} f_{n}(\mathbf{k})\to f_{n }(k)\exp\left[i\chi_{n}(\mathbf{k})\right].
\end{equation}
The closed loop at $k=\infty$ is therefore a pure gauge, which is necessarily a unit Berry phase,
\begin{equation}
\begin{split}
\exp\left[i\oint_{k=\infty} \mathbf{A}_n(\mathbf{k})\cdot d\mathbf{k}\right]&=\exp\left[i\chi_n|_0^{2\pi}\right]=1\\
&=\exp\left[i\int_{\mathbb{R}^2}  F_n(\mathbf{k})d^2\mathbf{k}\right].
\end{split}
\end{equation}
 $F_n(\mathbf{k})=\Hat{z}\cdot[\partial_\mathbf{k}\times\mathbf{A}_n(\mathbf{k})]$ is the Berry curvature and we have utilized Stokes' theorem to convert the line integral to a surface integral over the entire planar momentum space $\mathbb{R}^2$. Since the total Berry flux must come in multiples of $2\pi$, the Chern number $C_n$ is guaranteed to be an integer,
\begin{equation}
C_n=\frac{1}{2\pi}\int_{\mathbb{R}^2} F_n(\mathbf{k})d^2\mathbf{k}\in \mathbb{Z}.
\end{equation}
$C_n$ counts the number of singularities in the gauge potential $\mathbf{A}_n(\mathbf{k})$ as it evolves over the momentum space. We will now discuss the role of symmetries on the electromagnetic Chern number - specifically rotational symmetry.

\section{Rotational symmetry and spin}\label{app:PhotonSpin}

If the unit cell of the atomic crystal possesses a center (at least threefold cyclic) the response function is rotationally symmetric about $z$,
\begin{equation}
\mathcal{R}^{-1}\mathcal{M}(\omega,R\mathbf{k})\mathcal{R}=\mathcal{M}(\omega,\mathbf{k}).
\end{equation}
$R$ is the SO(2) matrix acting on the coordinates $\mathbf{k}$. $\mathcal{R}$ is the action of SO(2) acting on the fields $f$, which induces rotations in the $x$-$y$ plane,
\begin{equation}
R=\begin{bmatrix}
\cos\theta & \sin\theta\\
-\sin\theta & \cos\theta
\end{bmatrix},~~~~\mathcal{R}=\exp(i\theta\Hat{S}_z)=\begin{bmatrix}
R & 0\\
0 & 1
\end{bmatrix}.
\end{equation}
$\mathcal{R}$ is simply the SO(3) matrix along $\Hat{z}$ which rotates the polarization state of the electromagnetic field. Clearly the representation is single-valued (bosonic) and describes a spin-1 particle,
\begin{equation}
\mathcal{R}(2\pi)=\mathds{1}_3.
\end{equation}
Infinitesimal rotations on the coordinates $\mathbf{k}$ gives rise to the orbital angular momentum (OAM) $\Hat{L}_z=-i\partial_\phi$ while infinitesimal rotations on the polarization state gives rise to the spin angular momentum (SAM) $\Hat{S}_z=-i\epsilon_{ijz}$. Consequently, the total angular momentum (TAM) $\Hat{J}_z$ is conserved, at all frequencies and wave vectors,
\begin{equation}
[\Hat{J}_z,\mathcal{M}(\omega,\mathbf{k})]=0, \qquad \Hat{J}_z=\Hat{L}_z+\Hat{S}_z.
\end{equation}
This implies the electromagnetic field is a simultaneous eigenstate of $\Hat{J}_z$,
\begin{equation}
\Hat{J}_zf_{n\mathbf{k}}=j_n f_{n\mathbf{k}}, \qquad j_n\in\mathbb{Z}.
\end{equation}
$j_n$ is necessarily an integer for photons. Note though, $j_n$ is only uniquely defined up to a gauge since we can always add an arbitrary OAM to the state $f_{n\mathbf{k}}\to f_{n\mathbf{k}}\exp(il_n'\phi)$ such that $j_n\to j_n+l_n'$.

\subsection{Stationary (high-symmetry) points}

At an arbitrary momentum $\mathbf{k}$, the SAM and OAM are not good quantum numbers - only the TAM is well defined (up to a gauge). However, at stationary points $k=k_i$, also known as high-symmetry points (HSPs), the electromagnetic field is a simultaneous eigenstate of $\hat{S}_z$ and $\hat{L}_z$. In the continuum limit there are two such HSPs, $k_i=0$ and $k_i=\infty$. At these specific momenta, the response function is rotationally invariant - it commutes with $\mathcal{R}$,
\begin{equation}
[\mathcal{R},\mathcal{M}(\omega,k_i)]=[\hat{S}_z,\mathcal{M}(\omega,k_i)]=[\hat{L}_z,\mathcal{M}(\omega,k_i)]=0.
\end{equation}
Since $\mathcal{M}$ is a continuous function of $\mathbf{k}$, it cannot depend on the azimuthal coordinate $\phi$ at HSPs, otherwise $\mathcal{M}$ would be multivalued. Hence, the electromagnetic field is an eigenstate of both $\hat{S}_z$ and $\hat{L}_z$ at $k_i$,
\begin{equation}
\hat{S}_zf_{n\mathbf{k}_i}=m_n(k_i)f_{n\mathbf{k}_i},\qquad \hat{L}_zf_{n\mathbf{k}_i}=l_n(k_i)f_{n\mathbf{k}_i}.
\end{equation}
$m_n(k_i)=\pm 1,0$ is the SAM eigenvalue at $k_i$ of the $n$th band and $l_n(k_i)\in\mathbb{Z}$ is the OAM eigenvalue. Importantly, only the SAM is gauge invariant because it represents the eigenvalue of a matrix - ie. it only depends on the polarization state. This immediately implies the eigenmode can be factored into a spin and orbital part at HSPs,
\begin{equation}
f_{n\mathbf{k}_i}\propto [e_m(k_i)]_n \exp[il_n(k_i)\phi].
\end{equation}
$[e_m(k_i)]_n$ is the particular spin eigenstate at $k_i$ for the $n$th band. There are three possible eigenstates $e_m$ corresponding to three quantized spin-1 vectors,
\begin{equation}
\mathcal{R}e_m=e^{im\theta}e_m,\qquad \hat{S}_ze_m=me_m,
\end{equation}
where $m=\pm 1, 0$ labels the quantum of spin for each state,
\begin{equation}
e_{\pm }=\frac{1}{\sqrt{2}}\begin{bmatrix}
1 \\ \pm i \\ 0
\end{bmatrix}, \qquad e_0=\begin{bmatrix}
0 \\ 0\\ 1
\end{bmatrix}.
\end{equation}
$e_\pm$ are right and left-handed states respectively and represent electric resonances $E_y=\pm iE_x$ with $H_z=0$. The spin-0 state $e_0$ is a magnetic resonance $E_x=E_y=0$ with $H_z\neq 0$.

\subsection{Spin spectrum}

To determine the spin state of a particular band $n$, we need to solve the wave equation at HSPs. At these points, only three parameters are permitted by symmetry,
\begin{equation}
\mathcal{M}(\omega,k_i)=\begin{bmatrix}
\varepsilon & ig  & 0\\
-ig & \varepsilon & 0\\
0 & 0 & \mu
\end{bmatrix}.
\end{equation}
$\varepsilon$ and $\mu$ are the scalar permittivity and permeability respectively. $g$ is the gyrotropic coefficient which breaks both time-reversal ($\mathcal{T}$) and parity ($\mathcal{P}$) symmetry but preserves rotational ($\mathcal{R}$) symmetry. Assuming a regularized response function, nontrivial solutions of the wave equation simultaneously satisfy,
\begin{equation}\label{eq:Determinant}
\det[\mathcal{M}(\omega,k_i)]=0 ,\qquad \omega=\omega_n(k_i)\neq 0.
\end{equation}
There are three possible conditions that satisfy Eq.~(\ref{eq:Determinant}). The first two generate right or left-handed states $e_\pm$,
\begin{equation}
m_n(k_i)=\frac{g(\omega_{n}(k_i),k_i)}{\varepsilon(\omega_{n}(k_i),k_i)}=\pm 1.
\end{equation}
The last generates the the spin-0 state $e_0$,
\begin{equation}
m_n(k_i)=\mu(\omega_{n}(k_i),k_i)=0.
\end{equation}
Note, since $m_n$ is a discrete quantum number, it cannot vary continuously if rotational symmetry is preserved. It can only be changed at a topological phase transition which requires an accidental degeneracy at a HSP.

\subsection{Symmetry-protected topological (SPT) phases}\label{sec:SPTphases}

Remarkably, the electromagnetic Chern number is determined entirely from the spin eigenvalues at the HSPs $k_i$. The proof is surprisingly simple. Due to rotational symmetry, the Berry curvature $F_n(k)=\partial_k A_n^\phi(k)$ depends only on the magnitude of $k$ since $F_n$ is a scalar. Integrating the Berry curvature over all space $\mathbb{R}^2$, we arrive at,
\begin{equation}
C_n=A_n^\phi(\infty)-A_n^\phi(0)=l_n(\infty)-l_n(0).
\end{equation}
This follows because $f_{n\mathbf{k}_i}$ is an eigenstate of the OAM at HSPs $k_i=0$ and $k_i=\infty$. The OAM at $k_i$ is not gauge invariant, however the \textit{difference} at the two stationary points is gauge invariant because the TAM is conserved $j_n=m_n(0)+l_n(0)=m_n(\infty)+l_n(\infty)$. Substituting for $m_n$ we obtain,
\begin{equation}
C_n=m_n(0)-m_n(\infty).
\end{equation}
Hence, the spin eigenstate must change at HSPs $m_n(0)\neq m_n(\infty)$ to acquire a nontrivial phase $C_n\neq 0$. It is also clear that a purely gyrotropic medium $\mu=1$ always has Chern numbers of $|C_n|=2$ since $m_n(k_i)=\pm 1$ only assumes two values. $|C_n|=1$ is much more exotic as it requires both gyrotropy and magnetism.

\bibliography{natHyp2.bib}

\begin{thebibliography}{77}
\bibitem [{not()}]{note1}%
  \BibitemOpen
  \href@noop {} {\ }\bibinfo {note} {In the rare circumstance of strong
  magnetoelectricity $\chi_i\neq 0$, it is conceptually possible to break
  $\mathcal{T}$ symmetry without dispersion.}\BibitemShut {Stop}%
\end{thebibliography}

\end{document}